\documentclass{eds2019}

\def\Journal#1#2#3#4{{#1} {\bf #2}, #3 (#4)}

\def\PLB{{\em Phys. Lett.}  B}

\def\EPJC{{\em Eur. Phys. J.} C}

\def\be{\begin{equation}}
\def\ee{\end{equation}}
\def\bea{\begin{eqnarray}}
\def\eea{\end{eqnarray}}

\newcommand{\Photo}{}

\begin{document}
\vspace*{4cm}
\title{RECENT CMS AND CMS-TOTEM RESULTS ON \\ DIFFRACTION AND EXCLUSIVE PRODUCTION}

\author{O. SUR\'ANYI on behalf of the CMS Collaboration}

\address{MTA-ELTE Lend\"ulet CMS Particle and Nuclear Physics Group, \\ E\"otv\"os Lor\'and University, Budapest, Hungary; oliver.suranyi@cern.ch}

\maketitle\abstracts{
Exclusive and diffractive physics measurements are important for the better understanding of the non-perturbative regime of QCD. Two recent results of the CMS experiment is presented in this paper. The total and differential cross sections of central exclusive $\pi^+\pi^-$ production are measured at $5.02$ and $13$~TeV in the $p_{\mathrm{T}}(\pi) > 0.2~$GeV/$c$ and $|\eta(\pi)| < 2.4$ kinematic region. The invariant mass distribution is fitted by the sum of a continuum and four interfering relativistic Breit-Wigner functions. In the second part of the paper the measurement of the single diffractive dijets is presented, which are studied by using proton tagging capabilities of the TOTEM analysis. A joint CMS-TOTEM study is carried out and the total and differential cross sections are measured in the $0.03 < |t| < 1.0~$GeV$^2$ and $0 < \xi < 0.1$ kinematic region.}

\section{Introduction}
The field of diffractive and exclusive physics is an active research topic. Many important questions are studied, such as the properties of low mass scalar and tensor resonances, the structure of the pomeron and soft survival effects. This paper introduces two results from the CMS experiment: the measurement of central exclusive $\pi^+\pi^-$ production at $5.02$ and $13$~TeV~\cite{CMS:2019qjb} and single diffractive dijets with proton tagging at $8$~TeV~\cite{CMS:2018udy}.

\section{Central Exclusive $\pi^+\pi^-$ Production at 5.02 and 13~TeV}
\begin{figure}[t]
\centering
\includegraphics[width=0.7\textwidth]{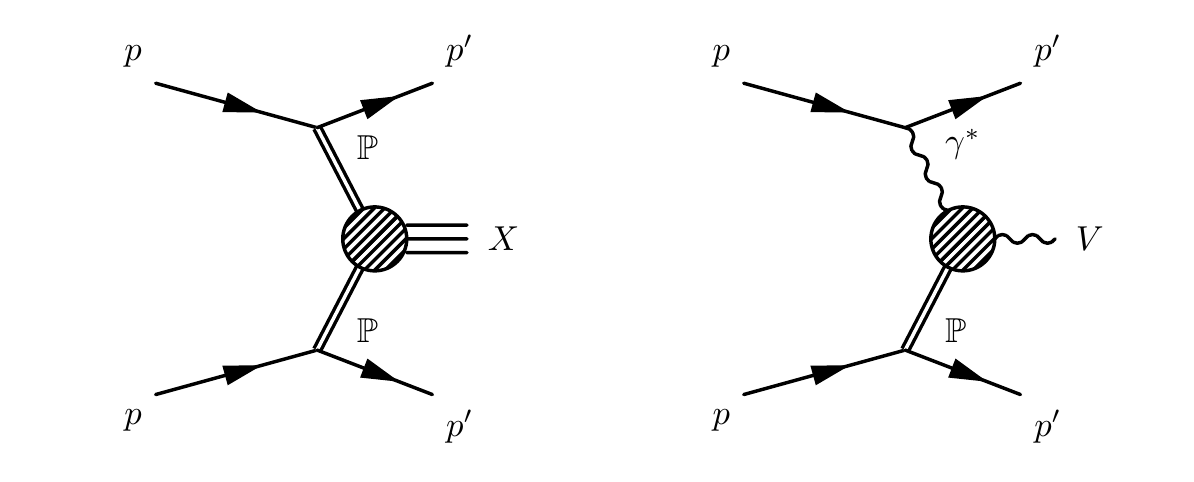}
\vspace{-5px}
\caption{The two dominant central exclusive production channels at LHC energies: double pomeron exchange (left) and vector meson photoproduction (right).~\protect\cite{CMS:2019qjb}}
\label{fig:cep}
\end{figure}

The central exclusive dihadron production process has been long used to study certain low mass resonances, as the quantum numbers of the exchanged photons and pomerons restrict the central system, practically working as a quantum number filter. The two dominant processes at LHC energies, double pomeron exchange (DPE) and vector meson photonproduction (VMP), are shown in Figure \ref{fig:cep}. In these types of collisions the beam protons remain intact and no other particles are produced beside the central system. When either one or both protons can dissociate diffractively, the process is called semi-exclusive production and it is considered a background in this measurement. This paper presents the cross section measurement of the central exclusive production with the $\pi^+\pi^-$ final state.

In the CMS, these processes are studied in collisions, which have low probability of multiple proton-proton collisions within the same bunch crossing (pileup). The data were collected with a zero bias trigger (random bunch crossings) in 2015 August and November with collision energies $13$~TeV and $5.02$~TeV respectively, with an average number of $0.3$--$0.5$ collisions per bunch crossings.

Selected events are required to have two oppositely charged tracks in the CMS tracking system. The difference in the $z$ coordinates of the closest approach to the beamline should be smaller than $3\sigma$ to ensure that the two tracks are originated from the same interaction. Pions are identified using the specific ionization ($\mathrm{d}E/\mathrm{d}x$) of the tracks in the tracking detectors. 
Furthermore, events must not contain any activity in the calorimeters, except in the $3\sigma$ region around the extrapolated track hits. The cross sections presented in this paper are measured in the $p_{\mathrm{T}}(\pi) > 0.2$~GeV/$c$ and $|\eta(\pi)| < 2.4$ kinematic region. All events are corrected for pileup rejection and tracking inefficiencies.


\begin{figure}[!b]
\centering
\includegraphics[width=0.44\textwidth]{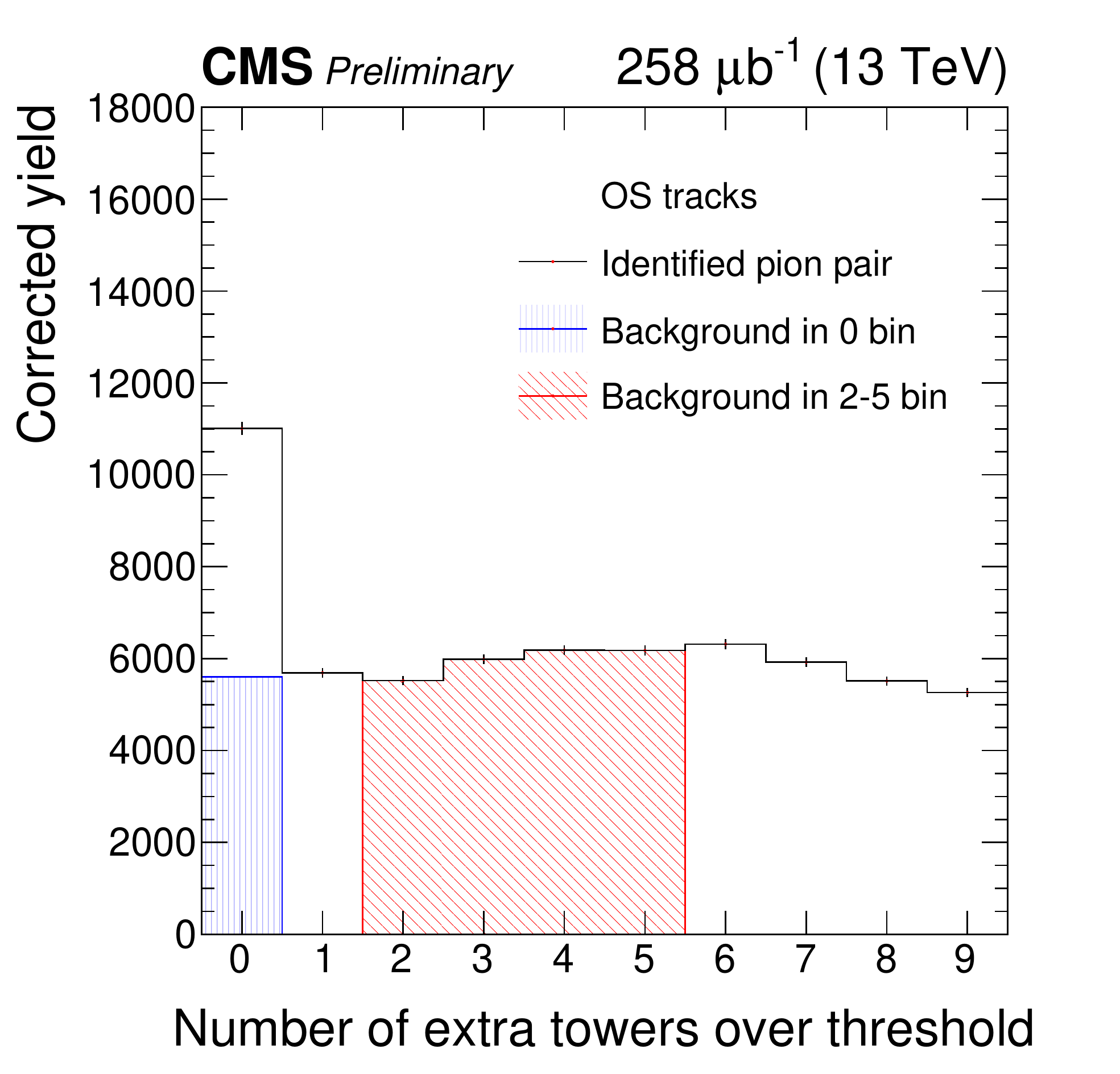}
\includegraphics[width=0.44\textwidth]{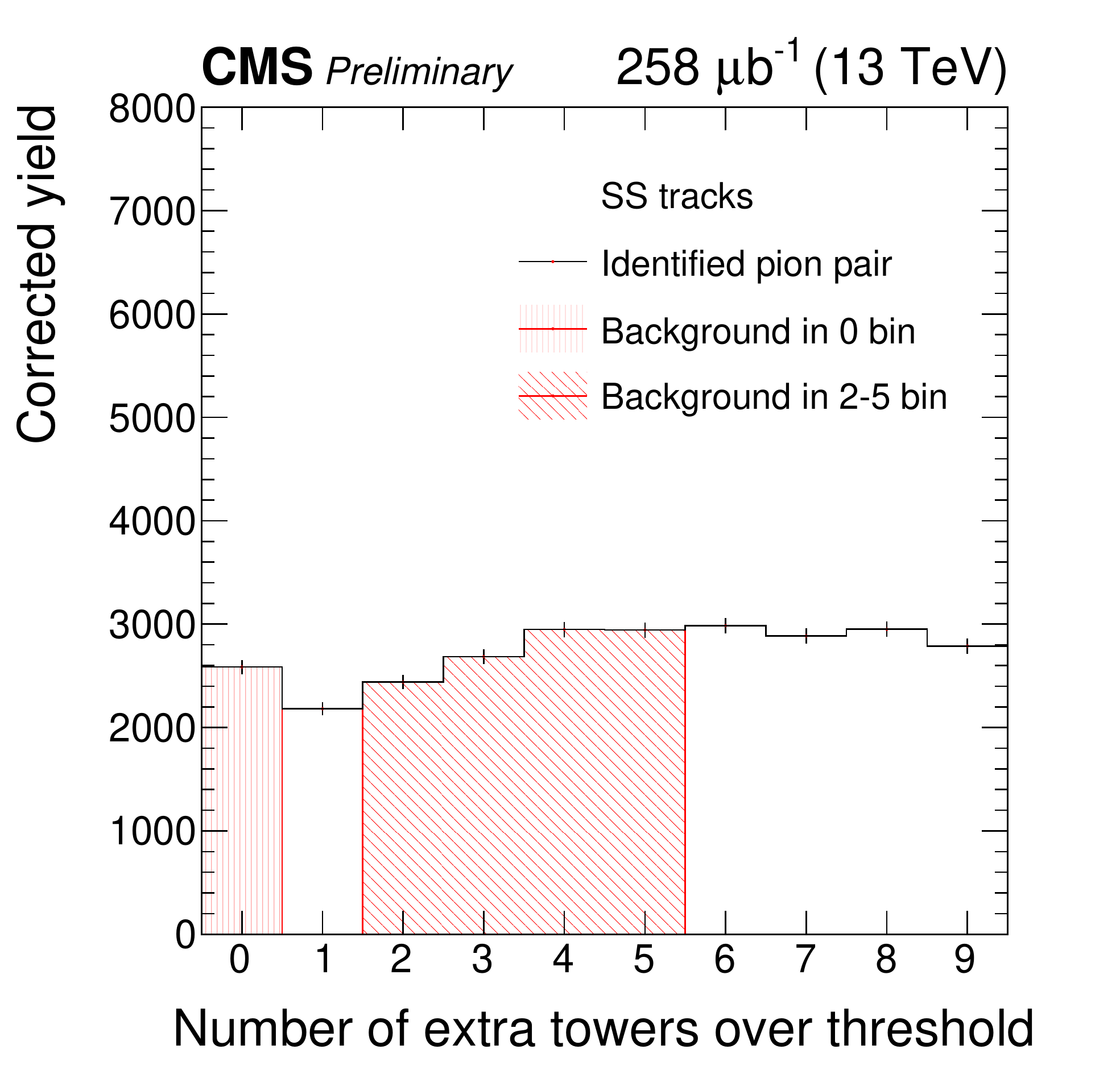}
\vspace{-5px}
\caption{The number of extra calorimeter hits in events with two opposite sign (left) and same sign (right) tracks.~\protect\cite{CMS:2019qjb}}
\label{fig:nonexc}
\end{figure}

The contamination of non-exclusive events gives the dominant background contribution. These events have some additional low-energy particles produced, which are not detected by the calorimeters. This background is estimated using a data-driven method, by using a data sample with extra calorimeter hits. The distribution of non-exclusive events are calculated by requiring $2$--$5$ calorimeter hits beside the ones corresponding to the pions. The normalization of this contribution is calculated from the calorimeter hit multiplicity distribution shown in Figure \ref{fig:nonexc}. The signal events are in the zero bin of the opposite sign sample. All events in the same sign sample are background events due to charge conservation. It is assumed that the ratio of the zero bin and $2$--$5$ extra hit region is the same for both opposite and same sign background events. The estimated background distribution is then subtracted from the final distributions.

\begin{figure}[!t]
\centering
\includegraphics[width=0.32\textwidth]{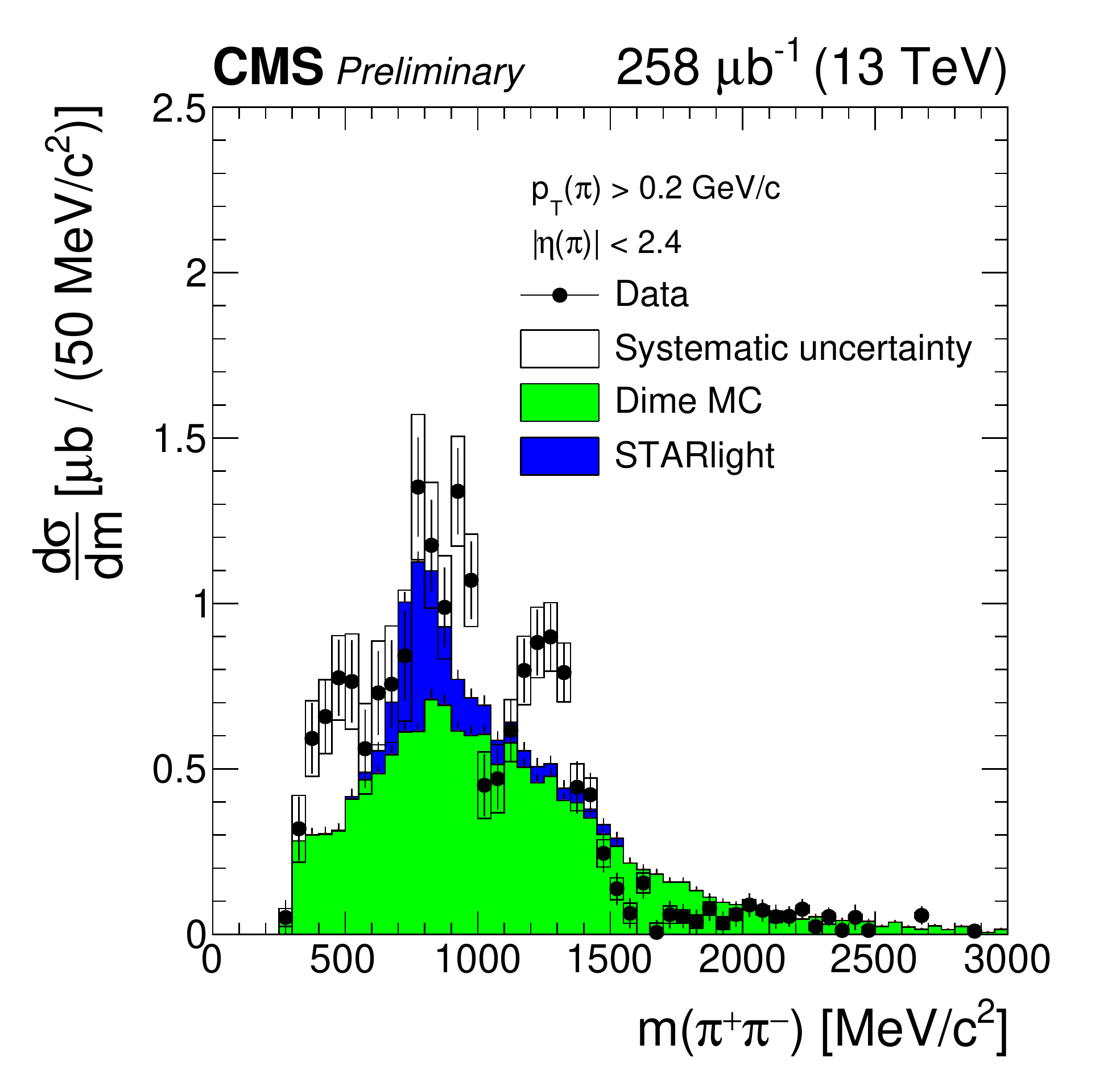}
\includegraphics[width=0.32\textwidth]{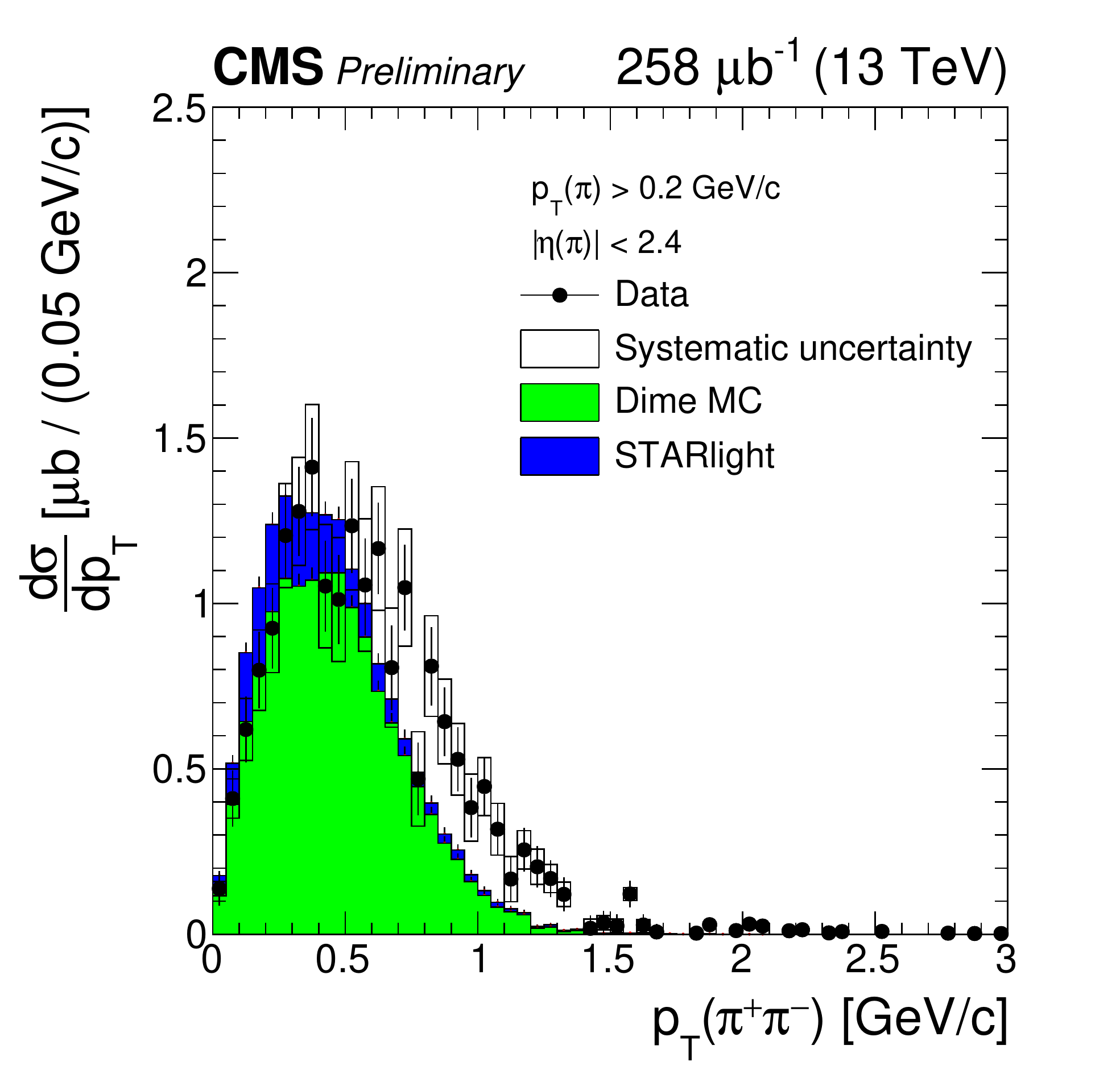}
\includegraphics[width=0.32\textwidth]{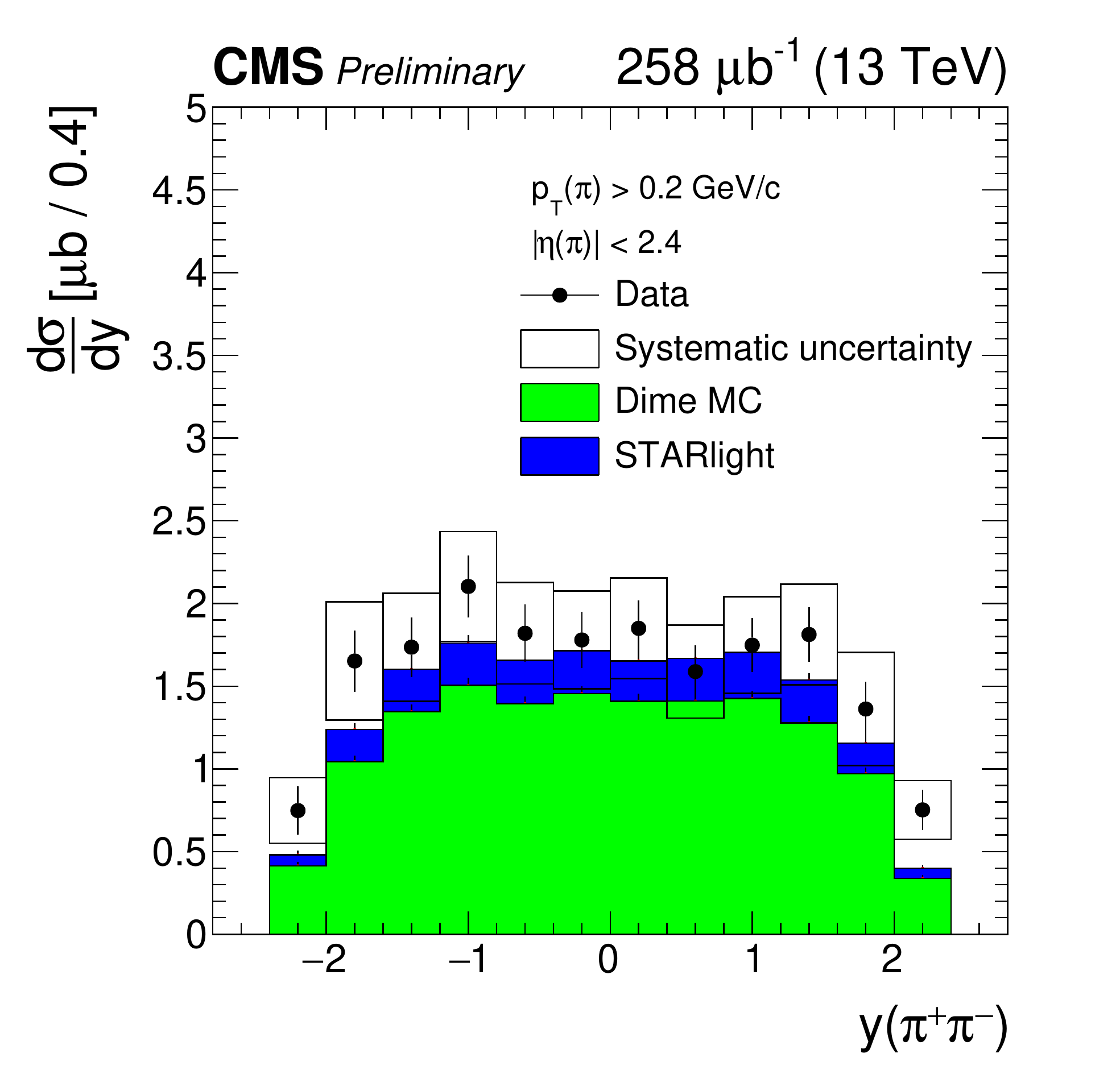}
\vspace{-5px}
\caption{The differential cross sections as a function of invariant mass (left), transverse momentum (middle) and rapitidity (right) of the pion pair.~\protect\cite{CMS:2019qjb}}
\label{fig:xsec}
\end{figure}

\begin{figure}[!b]
\centering
\includegraphics[width=0.44\textwidth]{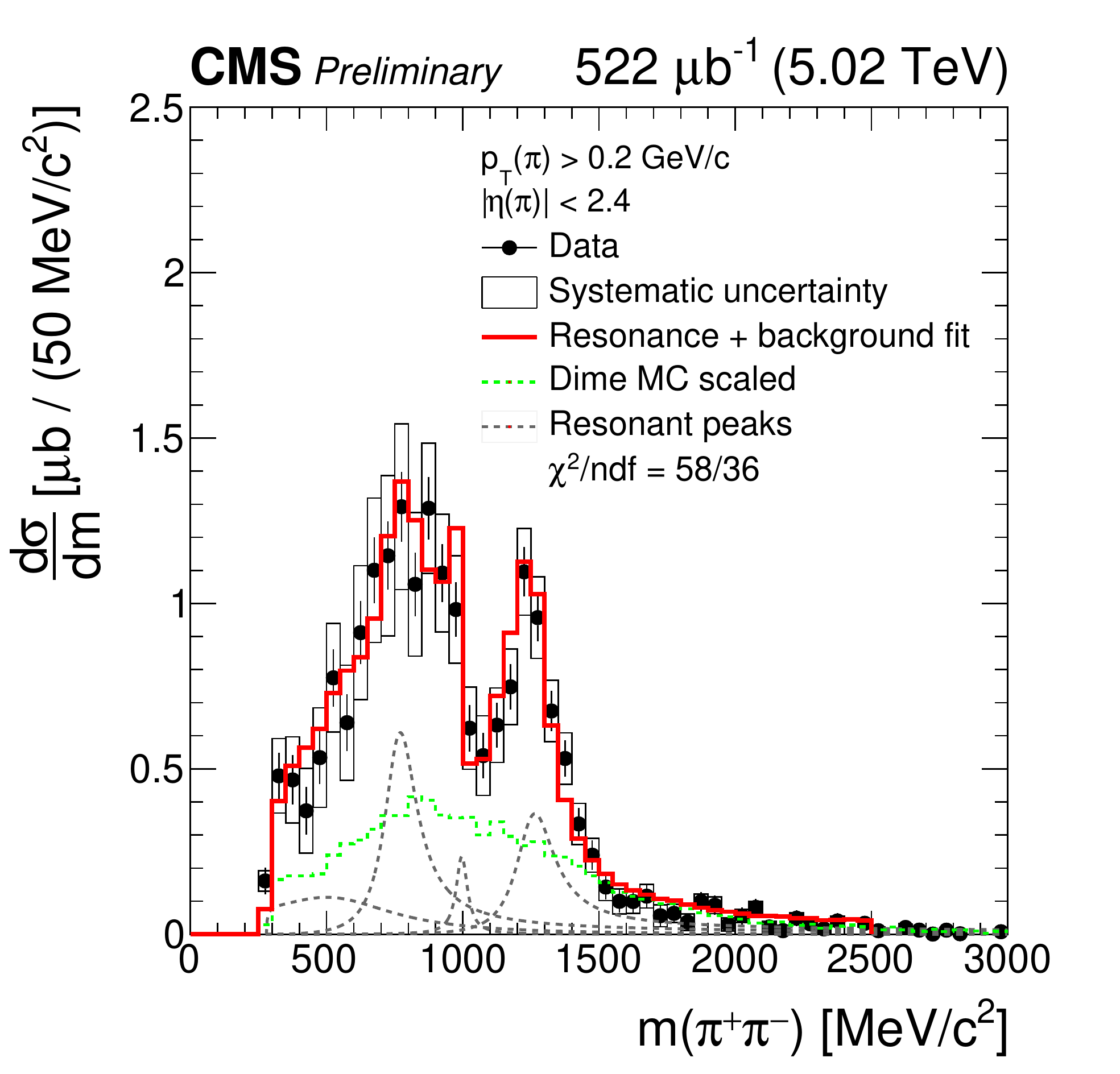}
\includegraphics[width=0.44\textwidth]{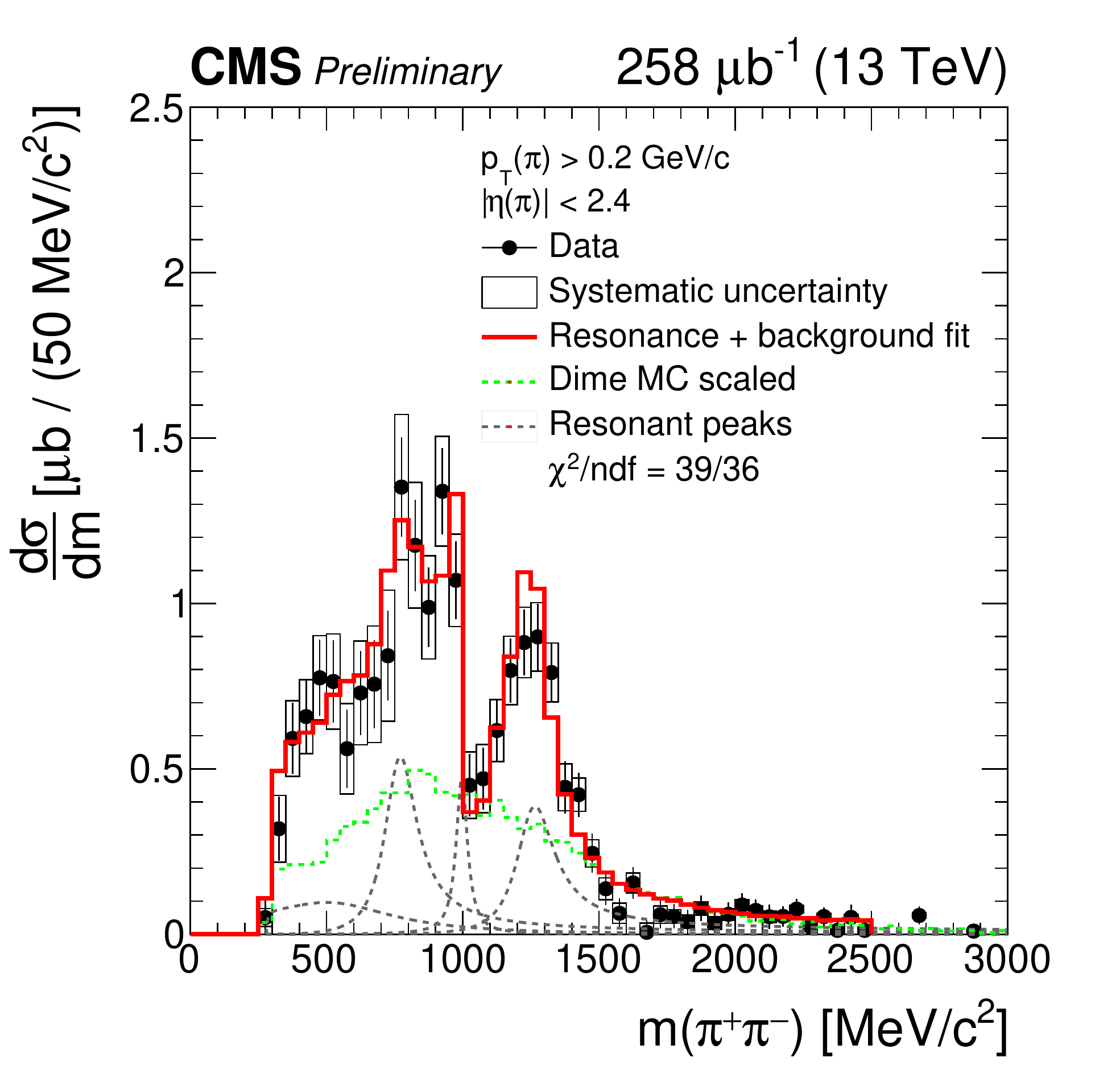}
\vspace{-5px}
\caption{The fit of invariant distributions with a sum of a continuum and four interfering relativistic Breit-Wigner functions.~\protect\cite{CMS:2019qjb}}
\label{fig:fit}
\end{figure}

The measured total cross section of this process is
\bea
\sigma(\sqrt{s} = 5.02~\mathrm{TeV}) &=& 19.6\!	\pm	0.4\!	~\mathrm{(stat.)} \pm	3.3\!	~\mathrm{(syst.)} \pm	0.01\!	~\mathrm{(lumi.)}~\mu \mathrm{b}, \\
\sigma(\sqrt{s} = 13~\mathrm{TeV}) &=& 19.0\!	\pm	0.6\!	~\mathrm{(stat.)} \pm	3.2\!	~\mathrm{(syst.)} \pm	0.01\!	~\mathrm{(lumi.)}~\mu \mathrm{b}.
\eea
The differential cross sections as a function of dipion invariant mass, transverse momentum and rapidity are shown in Figure \ref{fig:xsec}. These spectra are compared to Monte Carlo models: the \textsc{Dime MC} model~\cite{Harland-Lang:2013dia} which models non-resonant $\pi^+\pi^-$ production via DPE and the \textsc{STARlight} model~\cite{Klein:2016yzr} describing the $\rho^0 \to \pi^+\pi^-$ VMP channel. None of the existing Monte Carlo models describe the resonant DPE production yet. The mass distribution shows features previously seen at experiments with lower collision energy~\cite{Barberis:1999an}. There is a wide enhancement compared to predictions below $600~$MeV/$c^2$, which can be addressed to the presence of $f_0(500)$ resonance. At around $700$--$800$~MeV/$c^2$ there is a large peak, which is in agreement with the \textsc{STARlight} prediction. There is a sharp drop around $1000$~MeV/$c^2$, which can be interpreted as an interference effect between the $f_0(980)$ resonance and the continuum production channel as shown by the WA102 experiment~\cite{Barberis:1999an}. There is a prominent peak around $1250$~MeV/$c^2$ consistent with $f_2(1270)$ resonance. The \textsc{Dime MC} model overestimates the region around $1500$~MeV/$c^2$, which is probably due to the fact that this model was tuned on lower energy measurements.

The mass distribution is fitted with the sum of four interfering relativistic Breit-Wigner functions. The \textsc{Dime MC} distribution is included in the fit to describe the continuum production, but a scale factor is introduced to take into account the discrepancies. The whole fit function is convolved with a Gaussian to take into account the experimental resolution of the detector. The results of the fits are shown in Figure \ref{fig:fit}. 

\begin{figure}[!b]
\centering
\vspace{-5px}
\includegraphics[width=0.3\textwidth]{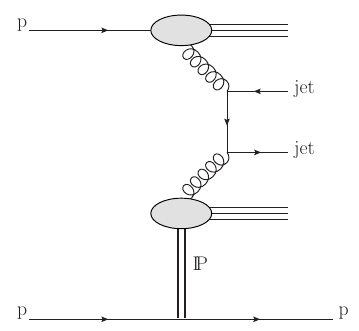}
\vspace{-5px}
\caption{Single diffractive dijet production.~\protect\cite{CMS:2018udy}}
\label{fig:dijet}
\end{figure}

\section{Single Diffractive Dijets with Proton Tagging at 8 TeV}
Diffractive processes give a significant contribution to the total cross section in proton-proton collisions. In a single diffractive collision one of the protons remains intact. This is characterized by a large rapidity gap around the proton. When a hard scale process is present, diffraction can be described by the convolution of diffractive parton distribution functions and hard scattering cross sections calculated from pQCD. However soft rescattering effects can break this factorization leading to the suppression of diffractive cross sections. The suppression factor is called rapidity gap survival probability. This paper reviews the measurement of single diffractive dijet cross section. The studied process is shown in Figure \ref{fig:dijet}.

\begin{figure}[!t]
\centering
\includegraphics[width=0.44\textwidth]{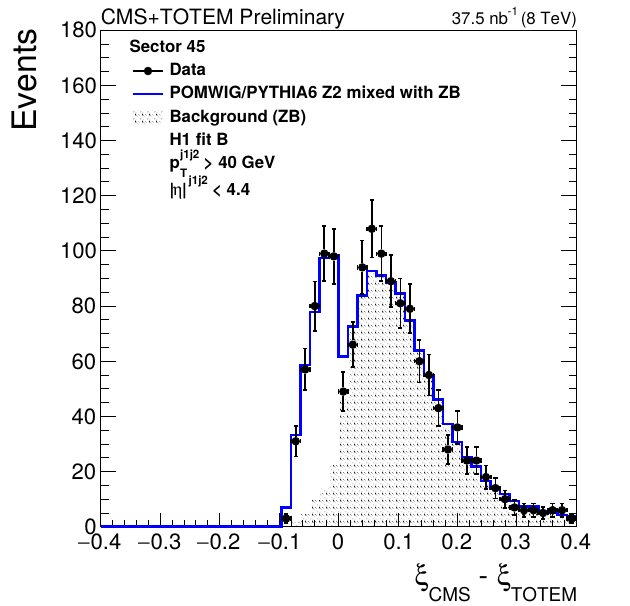}
\vspace{-5px}
\caption{Distribution of selected dijet and background events as a function of $\xi_{\mathrm{CMS}} - \xi_{\mathrm{TOTEM}}$.~\protect\cite{CMS:2018udy}}
\label{fig:dijet_bg}
\end{figure}

The scattered beam protons are detected by the Roman Pots (RP) of the TOTEM experiment~\cite{Anelli:2008zza}. These are small tracking detectors situated at $147$ and $220$~m distance from the interaction point on both side of the CMS. The $t$ squared four-momentum transfer is measured as
\be
t = (p_f-p_i)^2,
\ee
where $p_i$ and $p_f$ are the four-momentum of the proton before and after the collision respectively. The fractional momentum loss can be either measured by the RPs ($\xi_{\mathrm{TOTEM}}$) or estimated by using CMS information only ($\xi^{\pm}_{\mathrm{CMS}}$) as
\bea
\xi_{\mathrm{TOTEM}} &=& 1 - \frac{\left|\mathbf{p}_f\right|}{\left|\mathbf{p}_i\right|} \\
\xi^{\pm}_{\mathrm{CMS}} &=& \frac{\sum_i (E^i \pm p_z^i)}{\sqrt{s}},
\eea
where $\mathbf{p}_i$ and $\mathbf{p}_f$ are initial and final three-momentum of the beam proton, $\sqrt{s}$ is the center-of-mass energy, $E_i$ and $p_z$ is the energy and longitudinal momentum of reconstructed particles and the sum goes over all of the reconstructed particle candidates.

The measurement is performed using data with low pileup and special beam optics with $\beta^* = 90$~m collected in July 2012 with $8$~TeV collision energy. Dijet events are selected online by requiring two jets with $p_{\mathrm{T}} > 20$~GeV on the trigger level. Further offline event selection criteria are applied: the two jets are required to have $p_{\mathrm{T}} > 40$~GeV and $|\eta| < 4.4$, at least one reconstructed primary vertex and at least one RP with $0 < x_{\mathrm{RP}} < 7$~mm and $8.4 < y_{\mathrm{RP}} < 27$~mm. The cross sections are measured in the kinematic region $0.03 < |t| < 1.0$~GeV$^2$ and $0 < \xi_{\mathrm{TOTEM}} < 0.1$. The main source of the background is due to an overlap of a pp collision and an additional track in the RP. For signal events $\xi_{\mathrm{CMS}} < \xi_{\mathrm{TOTEM}}$ due to the limited pseudorapidity coverage of the CMS detector. Therefore the background is suppressed by requiring $\xi_{\mathrm{CMS}} - \xi_{\mathrm{TOTEM}} \leq 0$, as shown in Figure \ref{fig:dijet_bg}.

\begin{figure}[!b]
\centering
\includegraphics[width=0.98\textwidth]{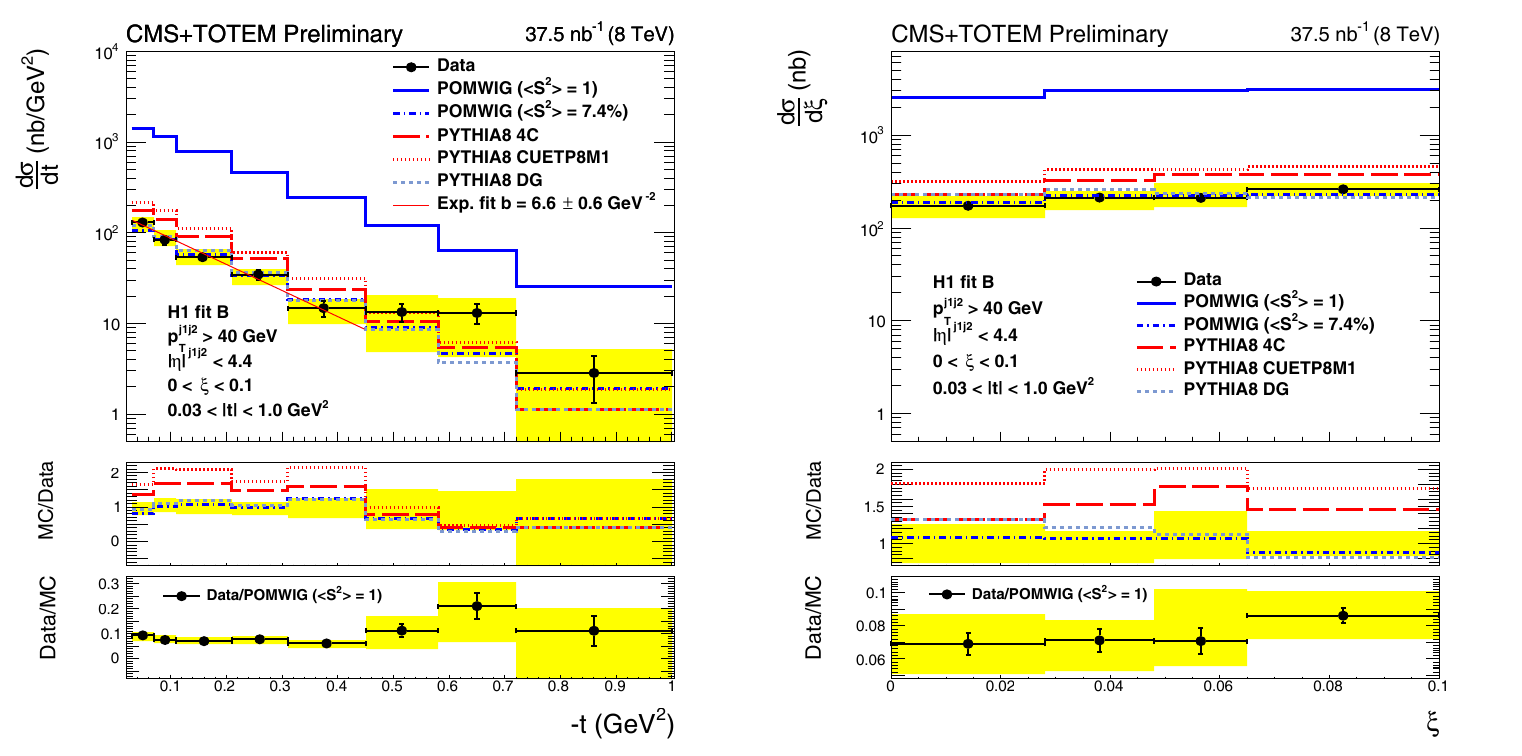}
\vspace{-5px}
\caption{Differential cross sections with respect to $-t$ and $\xi$.~\protect\cite{CMS:2018udy}}
\label{fig:dijet_res}
\end{figure}

\begin{figure}[!t]
\centering
\includegraphics[width=0.44\textwidth]{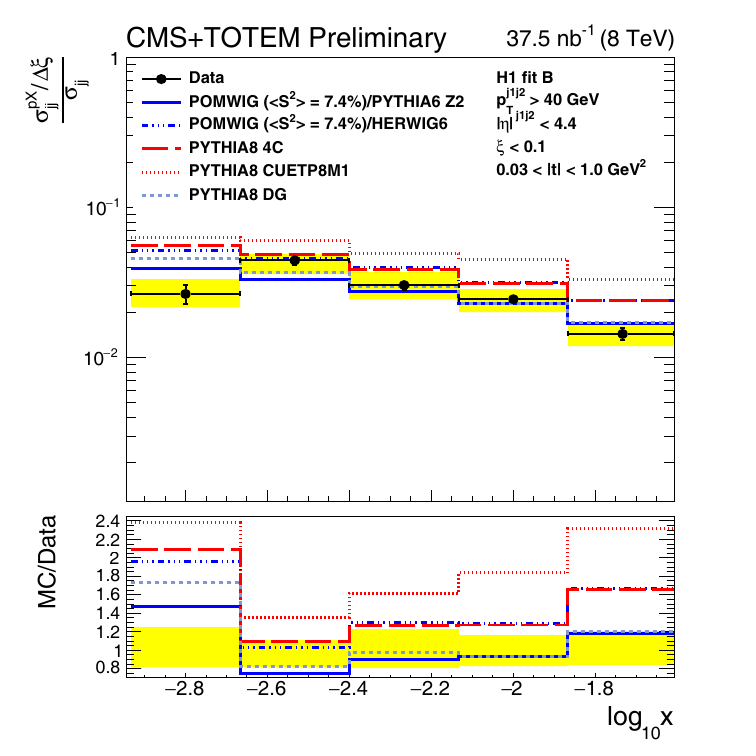}
\includegraphics[width=0.44\textwidth]{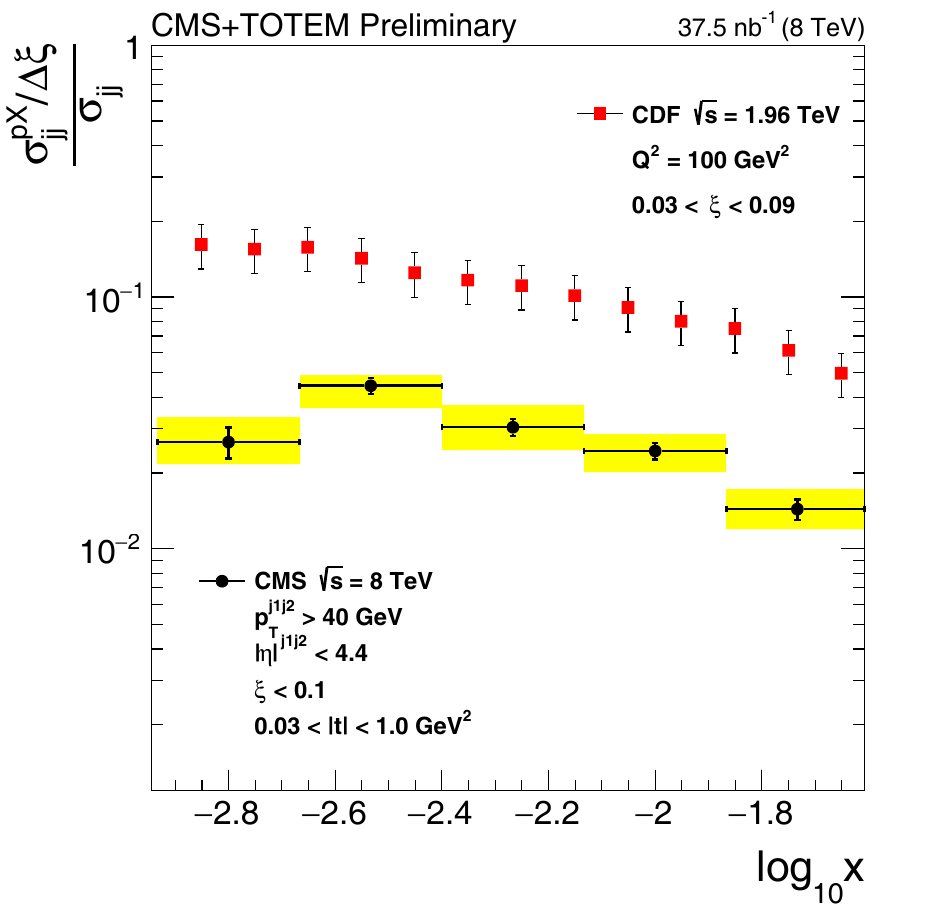}
\vspace{-5px}
\caption{Ratio of single diffractive and inclusive dijet production cross sections compared to MC predictions (left) and results from CDF (right).~\protect\cite{CMS:2018udy}}
\label{fig:dijet_res2}
\end{figure}

The total measured cross section of single diffractive dijet production in the kinematic region is
\be
\sigma_{jj}^{\mathrm{pX}} = 21.7 \pm 0.9~\mathrm{(stat)}~^{+3.0}_{-3.3}~\mathrm{(syst)} \pm 0.9~\mathrm{(lumi)}~\mathrm{nb}.
\ee
The differential cross sections are measured as a function of $-t$ and $\xi$, as shown in Figure \ref{fig:dijet_res}. The results are compared to \textsc{Pomwig}~\cite{Cox:2000jt} (with $100\%$ and $7.4\%$ gap survival probability) and \textsc{Pythia 8}~\cite{Sjostrand:2014zea} (with 4C, CUETP8M1 tunes and the Dynamic Gap (DG) model). In the DG model an event is classified as a diffractive process only when no additional partonic interactions occured. The measured cross sections are in agreement with \textsc{Pomwig} (with $7.4\%$ gap survival probability) and the \textsc{Pythia 8} DG model while, the \textsc{Pythia8} 4C and CUETP8M1 predicts a higher contribution. The ratio of the single diffractive dijet cross section and the inclusive dijet cross section is shown as a function of $\log_{10}x$ in the left panel of Figure \ref{fig:dijet_res2}. This ratio is also described better by the \textsc{Pomwig} and \textsc{Pythia 8} DG models. Finally, the cross section ratio is compared to the CDF result in the right panel of Figure \ref{fig:dijet_res2}. The cross section decreases compared to the one measured in CDF, which feature was already seen in CDF before.

\section{Conclusions}
The studying of diffractive and exlusive processes provides possibilities to study non-perturbative soft rescattering effects and the properties of pomeron and low mass resonances. The total and differential cross sections of exclusive $\pi^+\pi^-$ production were measured as a function of invariant mass, transverse momentum and rapidity of the pion pair. The mass spectrum is fitted with the sum of four interfering Breit-Wigner functions on the top of a continuum. The total and differential cross sections of single diffractive dijets production were also measured using the proton tagging capabilities of TOTEM Roman Pots.

\section*{Acknowledgments}
Supported by the \'UNKP-19-3 New National Excellence Program of the Ministry for Innovation and Technology, the National Research, Development and Innovation Office of Hungary (K~124845, K~128713, and K~128786) and the Hungarian Academy of Sciences ''Lend\"ulet'' (Momentum) Program (LP~2015-7/2015).

\end{document}